\documentclass[numberedappendix]{openjournal}

\usepackage{amsmath}

\usepackage{xcolor}
\usepackage{textgreek}
\usepackage[utf8]{inputenc}
\usepackage[english]{babel}

\usepackage{graphicx}

\usepackage{hyperref}
\hypersetup{
    unicode, 
    colorlinks=true,
    linkcolor=linkcolor,
    citecolor=linkcolor,
    filecolor=linkcolor,
    urlcolor=linkcolor,
}
\usepackage{color,colortbl}
\definecolor{linkcolor}{rgb}{0.0,0.3,0.5}
\usepackage{tensind}
\tensordelimiter{?}
\DeclareGraphicsExtensions{.bmp,.png,.jpg,.pdf}
\usepackage{verbatim}
\usepackage[normalem]{ulem}
\usepackage{orcidlink}
\usepackage{soul}
\definecolor{dgreen}{rgb}{0,0.6,0.0}

\newcommand{\be}{\begin{equation}}
\newcommand{\ee}{\end{equation}}
\newcommand{\beq}{\begin{equation*}}
\newcommand{\eeq}{\end{equation*}}
\newcommand{\bea}{\begin{eqnarray}}
\newcommand{\eea}{\end{eqnarray}}

\newcommand{\ba}{{\mathbf a}}
\newcommand{\bx}{{\mathbf x}}
\newcommand{\by}{{\mathbf y}}

\newcommand{\bn}{{\mathbf n}}
\newcommand{\bk}{{\mathbf k}}

\renewcommand{\bv}{{\mathbf v}}
\newcommand{\br}{{\mathbf r}}

\newcommand{\bnabla}{{\boldsymbol{\nabla}}}

\newcommand{\dd}{\partial}

\newcommand{\HH}{{\cal H}}

\newcommand{\de}{\delta}
\newcommand{\De}{\Delta}

\renewcommand{\th}{\theta}

\newcommand{\La}{\Lambda}

\newcommand{\si}{\sigma}

\newcommand{\om}{\omega}
\newcommand{\Om}{\Omega}

\newcommand{\ident}{{\rm 1\kern -2.5pt I}}

\newcommand{\cd}{\cdot}
\newcommand{\ra}{\rightarrow}

\begin{document}
\title{nonlinear velocity power spectrum: modeling the cosmological dependence on the Hubble constant and cold dark matter density}

\author{Francesca Lepori \orcidlink{0009-0000-5061-7138}}
\email{francesca.lepori@unige.ch}
\affiliation{D\'epartement de Physique Th\'eorique, Universit\'e de Gen\`eve, 24 quai Ernest-Ansermet, 1211~Gen\`eve~4, Switzerland}

\author{Ruth Durrer \orcidlink{0000-0001-9833-2086}}
\email{ruth.durrer@unige.ch}
\affiliation{D\'epartement de Physique Th\'eorique, Universit\'e de Gen\`eve, 24 quai Ernest-Ansermet, 1211~Gen\`eve~4, Switzerland}

\author{Martin Kunz \orcidlink{0000-0002-3052-7394}}
\email{martin.kunz@unige.ch}
\affiliation{D\'epartement de Physique Th\'eorique, Universit\'e de Gen\`eve, 24 quai Ernest-Ansermet, 1211~Gen\`eve~4, Switzerland}

\author{Francesco Sorrenti \orcidlink{0000-0001-7141-9659}}
\email{francescosorrenti@icc.ub.edu}
\affiliation{Institut de Ciències del Cosmos, Universitat de Barcelona (ICCUB), Martí i Franquès, 1, 08028 Barcelona, Spain}

\author{Julian Adamek \orcidlink{0000-0002-0723-6740}}
\email{adamekj@ethz.ch}
\affiliation{D\'epartement de Physique Th\'eorique, Universit\'e de Gen\`eve, 24 quai Ernest-Ansermet, 1211~Gen\`eve~4, Switzerland}
\affiliation{Institut f\"ur Teilchen- und Astrophysik, ETH Z\"urich, Wolfgang-Pauli-Strasse 27, 8093 Z\"urich, Switzerland}

\author{Davide Piras \orcidlink{0000-0002-9836-2661}}
\email{dpiras@ethz.ch}
\affiliation{D\'epartement de Physique Th\'eorique, Universit\'e de Gen\`eve, 24 quai Ernest-Ansermet, 1211~Gen\`eve~4, Switzerland}
\affiliation{Institut f\"ur Teilchen- und Astrophysik, ETH Z\"urich, Wolfgang-Pauli-Strasse 27, 8093 Z\"urich, Switzerland}

\begin{abstract}
In this paper we present a semi-analytical model for the velocity power spectrum in  $\La$CDM cosmology for wave numbers $k<1/$Mpc. We mainly concentrate on the dominant divergence part but also present some results on the vorticity contribution. We divide cosmological parameters into evolution and shape parameters and model the dependence of the evolution parameter $h$ and of the shape parameter $\om_{\rm cdm}$ with an accuracy better than 2.5\%. A surprising finding of our study is that the velocity power spectrum becomes independent of $\om_{\rm cdm}$ on nonlinear scales. A python implementation of the model is publicly available. \vspace{0.2cm}
\end{abstract}


\maketitle

\section{Introduction}
\label{sec:intro}

The statistical characterization of the large-scale structure of the Universe provides one of the main approaches to testing cosmological models. To this end, it is common to use summary statistics like the power spectrum or higher-order correlation functions of the galaxy distribution to compress the information available in the full density field. In particular, for a Gaussian random field, the power spectrum provides a complete statistical description and contains all the information about the fluctuations (see e.g.,~\citealp{Durrer:2020fza}).

At early times, fluctuations are small and can be approximated by linear perturbation theory. However, at late times, on intermediate-to-small scales, nonlinearities become important, and numerical simulations are required to model the distribution of matter in the Universe. 
In order to avoid running too many costly $N$-body simulations, different approaches have been developed to parametrize the nonlinear matter or galaxy power spectrum in terms of a limited set of functions calibrated on simulations (see e.g.,~\citealp{Smith:2002dz,Takahashi_2012,Mead:2015yca,Baumann_2012}).
More recently, \cite{Sanchez:2021plj} and \cite{Esposito:2024qlo} introduced the concept of evolution and shape parameters
to capture the response of the nonlinear power spectrum to variations in cosmological parameters.

A further complication arises from the fact that observations do not directly probe the total matter density field, but rather the distribution of luminous tracers such as galaxies. These tracers are biased with respect to the underlying matter density, introducing additional modeling uncertainties. In this work we instead focus on the velocity field, which is expected to be significantly less affected by bias than the galaxy density field \citep{Summers:1995ik,Zheng:2014vla}. Moreover, velocities are closely connected to the growth rate of structure, making them a powerful probe of cosmological evolution and gravity.

However, extracting velocity information from observations is considerably more challenging. Current approaches rely primarily on redshift-space distortions (RSD), which provide indirect access to velocity statistics and are one of the primary goals of present surveys \citep{Ross:2014qpa,eBOSS:2020yzd,eBOSS:2020uxp,
Euclid:2024yrr,DESI:2023dwi} as they provide a sensitive test of gravity. More direct reconstructions are based on peculiar velocity measurements such as Cosmicflows-4~\citep{Tully:2022rbj}. In addition, peculiar velocities leave imprints on a wide range of observables. For example, they contribute to the observed redshifts of type Ia supernovae, offering an alternative avenue to constrain cosmological parameters~\citep{Piras:2025gnx}.

These considerations motivate the development of fast and accurate models for the nonlinear velocity power spectrum that can be used in cosmological inference without requiring an excessive number of computationally expensive simulations.
In this paper, we develop a model for the cosmological velocity power spectrum based on the evolution mapping framework, which we implement in a publicly available code.\footnote{\url{https://github.com/leporif/velocity_power_spectrum}} We restrict our study to variations in the Hubble parameter $h$ and the matter density parameter $\Omega_\textrm{m}$, as motivated by an extension of the analysis of \cite{Piras:2025gnx}, which will be presented in an upcoming work. However, the approach presented here is more generally applicable.

Another study of the peculiar velocity power spectrum using evolution mapping was recently presented in~\cite{Esposito:2026zcm}. That work concentrates on evolution parameters only, albeit in a relatively wide class of models. Here, we restrict ourselves to flat $\La$CDM while considering not only evolution parameters, but also the shape parameter $\om_{\rm cdm}=\Omega_{\rm cdm}h^2$, where $\Omega_{\rm cdm}$ is the cold dark matter density. In this sense, the present work is complementary to~\cite{Esposito:2026zcm}. We do not study the shape parameter $\om_{\rm b}=\Omega_{\rm b}h^2$ (the baryon density) as it is very well measured by the cosmic microwave background (CMB) alone with little degeneracy with other parameters. Previous studies that model the nonlinear velocity power spectrum, e.g.,~\cite{Jennings:2010uv,Jennings:2012ej,Bel:2018awq,Koda:2013eya},
do not explicitly exploit the separation between evolution and shape parameters and consequently do not directly provide the $h$-dependence required for our application.

We note that all the measurements mentioned above are sensitive only to the radial velocity component. If the velocity field is a gradient field, this is sufficient to reconstruct the full velocity field. But if vorticity cannot be neglected, this is no longer sufficient. However, as we show in Appendix~\ref{app}, under the assumption of statistical homogeneity and isotropy the radial velocity field {\em is} sufficient for measuring both power spectra. In this case, the gradient and the vorticity power spectra can in principle be estimated from measurements of the radial velocity field alone. Furthermore, it is shown in~\cite{Jelic-Cizmek:2018gdp} that vorticity becomes important only at scales $k > 1 \, \mathrm{Mpc}^{-1}$, where, at redshift $z=0$, it starts to dominate over the gradient velocity power spectrum. For this reason, we focus our nonlinear modeling on scales $k \lesssim 1 \, \mathrm{Mpc}^{-1}$, where vorticity is still a small contribution.

The structure of the paper is as follows. In Section~\ref{sec:meth}, we recall the evolution mapping approach. In Section~\ref{sec:evol-h}, 
we model the deviations from the exact mapping of Eq.~\eqref{eq:evol} in the 
nonlinear regime and validate the evolution mapping approach in the simplified 
case of a single evolution parameter, $h$. In Section~\ref{sec:shape-cdm}, we model 
the dependence of the velocity power spectrum on the shape parameter 
$\omega_{\rm cdm}$. Finally, in Section~\ref{sec:full-model}, we present our complete 
model incorporating the dependence on both $h$ and $\omega_{\rm cdm}$. Our conclusions are given in Section~\ref{sec:concl}. Additional material is presented in the Appendices. In Appendix~\ref{ap:vort}, we introduce a model for the vorticity power spectrum and discuss its contribution. In Appendix~\ref{app} we prove the remarkable result that under the assumption of statistical homogeneity and isotropy, from measurements of the radial component of the velocity field alone, one can actually infer its gradient and vorticity correlation functions and power spectra.

\section{Evolution mapping}
\label{sec:meth}
We develop our model for the nonlinear velocity power spectrum based on the 
evolution mapping approach developed in ~\cite{Sanchez:2021plj}. In this framework, 
cosmological parameters are classified as either evolution parameters or shape 
parameters. Evolution parameters affect only the amplitude of the linear matter power spectrum at a given redshift, while shape parameters determine its shape. 
Although the evolution mapping approach was originally developed for the matter power spectrum, here we shall apply it to the velocity-divergence power spectrum. In linear theory the power spectra of the velocity divergence $\theta$ and of matter density fluctuations are directly related through
\begin{equation}
P_{\theta\theta}(k, z) = f^2(z) \mathcal{H}^2(z)\, P_L(k, z) \ ,
\end{equation}
where $f(z) = d\ln D/d\ln a$ is the linear growth rate, $D$ is the growth factor, $a$ is the scale factor, $\mathcal{H}(z)=aH(z)$ is the conformal Hubble parameter and $P_L$ is the linear matter density power spectrum. Since the evolution mapping approach applies equally to 
$P_{\theta\theta}$, in this work we develop a model for the normalized velocity power spectrum,
\begin{equation}
{P}(k, z) := \frac{P_{\theta\theta}(k, z)}{f^2(z) \HH^2(z)} \ ,
\end{equation}
which reduces to the matter power spectrum in linear theory. Note that the $h$ and $\om_{\rm cdm}$ dependence of $P(k,z)$ is different from the dependence of $P_{\theta\theta}(k, z)$ since the prefactor $1/(f^2(z) \HH^2(z))$ also depends on both these parameters.
In~\cite{Sanchez:2021plj} it is shown that the Hubble parameter $h$ is an evolution parameter. To see this, it is however important not to have hidden dependences on $h$ e.g.,~in the length scales. We therefore measure comoving lengths (and wave numbers) in Mpc (respectively Mpc$^{-1}$) without a factor $h^{-1}$ (respectively $h$).

The amplitude of the linear power spectrum can be conveniently parametrized by $\sigma_{12}$, defined as
\begin{equation}
\sigma_{12}^2(z) = \int_0^{\infty} \frac{k^2\,dk}{2\pi^2}\, P_L(k, z) \left| W(kR) 
\right|^2 \, .
\end{equation}
Here $R=12$ Mpc and $W(kR)$ is the Fourier transform of the top-hat window function in real space,
\begin{equation}
W(kR) = \frac{3}{(kR)^3} \left[ \sin(kR) - kR \cos(kR) \right] = \frac{3}{kR}\,j_1(kR) \,,
\end{equation}
where $j_n$ denotes the spherical Bessel function of order $n$. We use $\si_{12}$, the variance at scale $R=12\,\mathrm{Mpc}$, instead of the more common $\si_8$ where $R=8\,h^{-1}\,\mathrm{Mpc}$, due to our convention not to include factors of $h$ in units, as mentioned above.
In addition, this is the variance of the total matter perturbation, not only the CDM (see~\citealp{Castorina:2015bma}).

In linear perturbation theory, evolution parameters are perfectly degenerate with 
$\sigma_{12}$. This means that varying an evolution parameter while keeping 
all shape parameters fixed leaves the shape of the matter power spectrum 
unchanged, only rescaling its amplitude. As a consequence, the matter power 
spectrum of any new cosmology can be recovered from a fiducial spectrum by simply 
finding the redshift $z'$ at which both cosmologies share the same $\sigma_{12}$. 
Specifically, for any cosmology $c$, there exists a redshift $z'$ such that
\begin{equation}
P_{L}^{\rm fid}(k, z') = P_{L}^{c}(k, z), \label{eq:evol}
\end{equation}
where $z'$ is determined by the condition $\sigma_{12}^{\rm fid}(z') = 
\sigma_{12}^{c}(z)$.
This redshift mapping eliminates the need to compute the matter power spectrum 
for each value of the evolution parameters.
Beyond linear theory, this mapping is no longer exact, as nonlinear 
gravitational evolution introduces additional cosmological dependence that 
cannot be absorbed into $\sigma_{12}$ alone.

\section{Nonlinear evolution mapping for the Hubble parameter}
\label{sec:evol-h}

As mentioned, at fixed $\omega_{\rm cdm}$ and $\omega_{\rm b}$, the Hubble parameter $h$ is an evolution parameter. In this section, we test the validity of the evolution mapping approach for $h$ in the nonlinear regime. 
In order to test the validity of this mapping beyond linear theory,  we run $N$-body simulations with \texttt{gevolution}~\citep[][]{Adamek:2015eda,Adamek:2016zes} \footnote{For details and updates on \texttt{gevolution} see also \url{https://gevolution-code.net/}.} with different values of $\{h, A_{\rm s}\}$, where $A_s$ is the amplitude of scalar curvature perturbations (see, e.g.~\citealp[][Chapter 3]{Durrer:2020fza} for a definition). For this test we choose $A_{\rm s}$ in such a way that all these models have the same $\sigma_{12}$ at $z = 0$.

\begingroup 
    \setlength{\tabcolsep}{10pt} 
    \renewcommand{\arraystretch}{1.5} 
    \setlength\extrarowheight{2pt}
    \begin{table}
        \centering
        \begin{tabular}{ c c c c c c}  
 $h$ & $A_{\rm s}$ & $\sigma_{12} = 0.5111$ & $\sigma_{12} = 0.6473$ & $\sigma_{12} = 0.7176$ & $\sigma_{12} = 0.8409$ \\
\hline
  0.67556 & $2.215\times10^{-9}$ & 1 & 0.5 & 0.3 & 0 \\
 0.47556 & $1.6349\times10^{-9}$ & 0.7649  & 0.3658 & 0.2138 & 0 \\
 0.57556 & $1.9188\times10^{-9}$ & 0.8837 & 0.4332 & 0.2569 & 0 \\
  0.77556 & $2.5229\times10^{-9}$ & 1.1139 & 0.5662 & 0.3432 & 0 \\
  0.87556 & $2.8414\times10^{-9}$ & 1.2255 & 0.6316 & 0.3862 & 0 \\
        \end{tabular}  
\caption{Redshifts at which $\sigma_{12}$ reaches the specified values for the different $(h,A_{\rm s})$ models considered in Figs.~\ref{fig:delta_panels} and \ref{fig:theta_panels}. All simulations have a box size of $L_{\mathrm{box}}\simeq 190\,\mathrm{Mpc}$, contain $1024^3$ particles, and use $512^3$ grid points. The first row corresponds to the fiducial model. To estimate the derivative correction, we run an additional set of simulations following the same setup but adopting a finer spacing in $h$, $\Delta h = 0.025\,h_{\rm fid}$, as described in the text.}
        \label{table:h-sims}
    \end{table}
\endgroup

In Table~\ref{table:h-sims}, we report the values of the cosmological parameters used in the simulation runs (columns $1$ and $2$), and the redshift at which $\sigma_{12}$ reaches specified values for the different models (columns from $3$ to $6$). 
The other cosmological parameters are held fixed and correspond to our fiducial model which was motivated by the best fit Planck 2013 cosmology~\citep{Planck:2013pxb},
\be
h = 0.67556\,,\quad \om_{\rm cdm} = 0.12038  \,, \quad \quad \om_{\rm b} = 0.022032\,, \quad A_\mathrm{s} = 2.215 \times 10^{-9}\,,\quad n_{\rm s} =0.9619 \,.
\ee

Since, in linear perturbation theory, evolution parameters do not affect the spectrum evaluated at the same value of $\si_{12}$, we can hope that their effect on the nonlinear power spectrum is also relatively modest and can be modeled by a first-order Taylor series. Taking into account only $h$, we therefore model its dependence via
\begin{equation}
P(k, \sigma_{12}, h) \approx P(k, \sigma_{12}, h_{\rm fid})
+ \frac{\partial P(k, \sigma_{12}, h)}{\partial h}\big|_{h_{\rm fid}}
(h-h_{\rm fid}). \label{eq:h-nl-corr}
\end{equation}

We estimate the derivative of the power spectrum with respect to $h$ in the nonlinear regime using numerical simulations \footnote{In the linear regime, this derivative at fixed $\sigma_{12}$ is identically zero.}. Specifically, we employ a four-point, fourth-order central finite-difference stencil, which requires four additional simulations with perturbed values of $h$. We adopt a step size of $\Delta h = 0.025\,h_{\rm fid}$.

\begin{figure}[htbp]
    \centering
\begin{minipage}{0.45\textwidth}
    \centering
    \includegraphics[width=\linewidth]{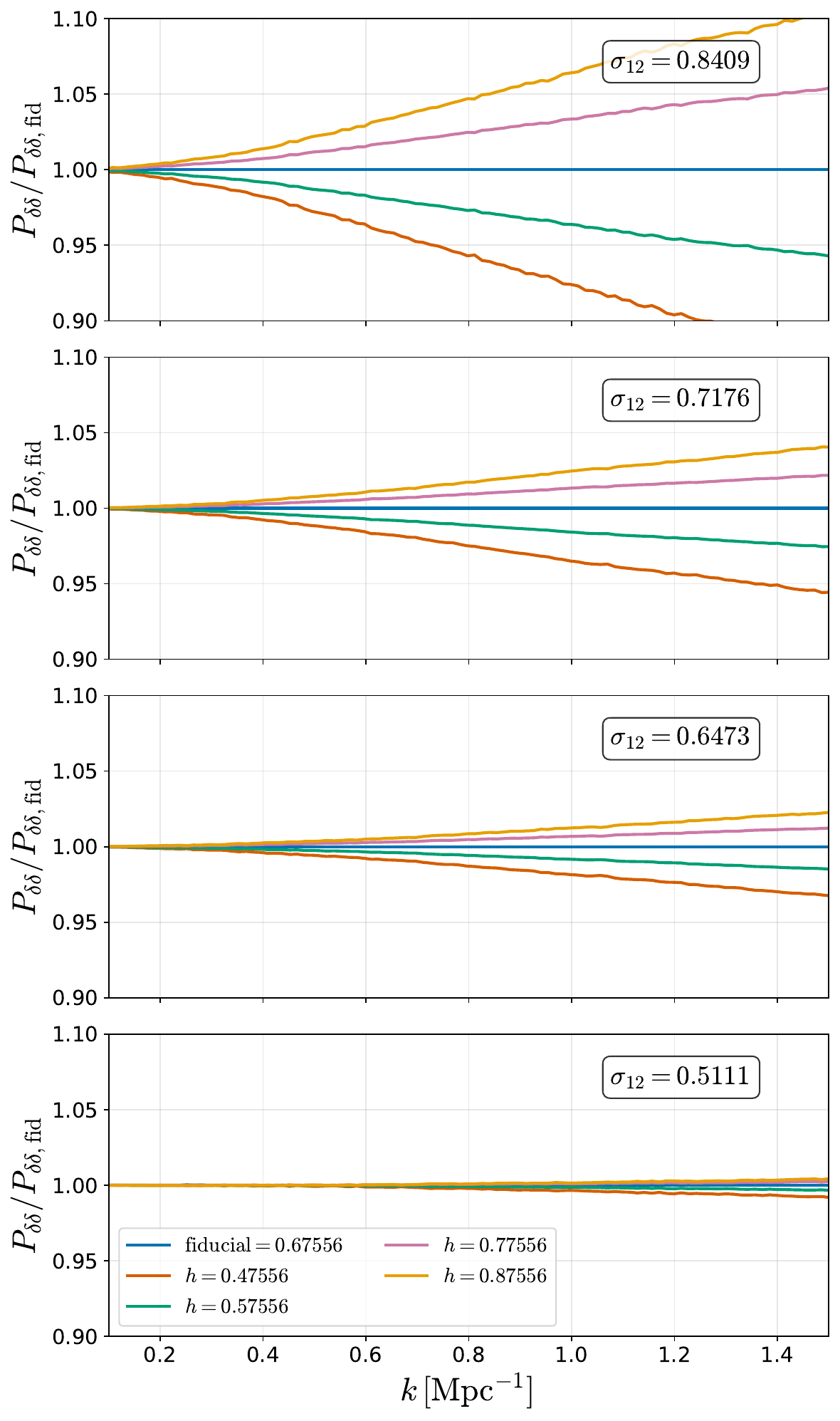}\\
    {\small (a) Without nonlinear correction.}
\end{minipage}
\hfill
\begin{minipage}{0.45\textwidth}
    \centering
    \includegraphics[width=\linewidth]{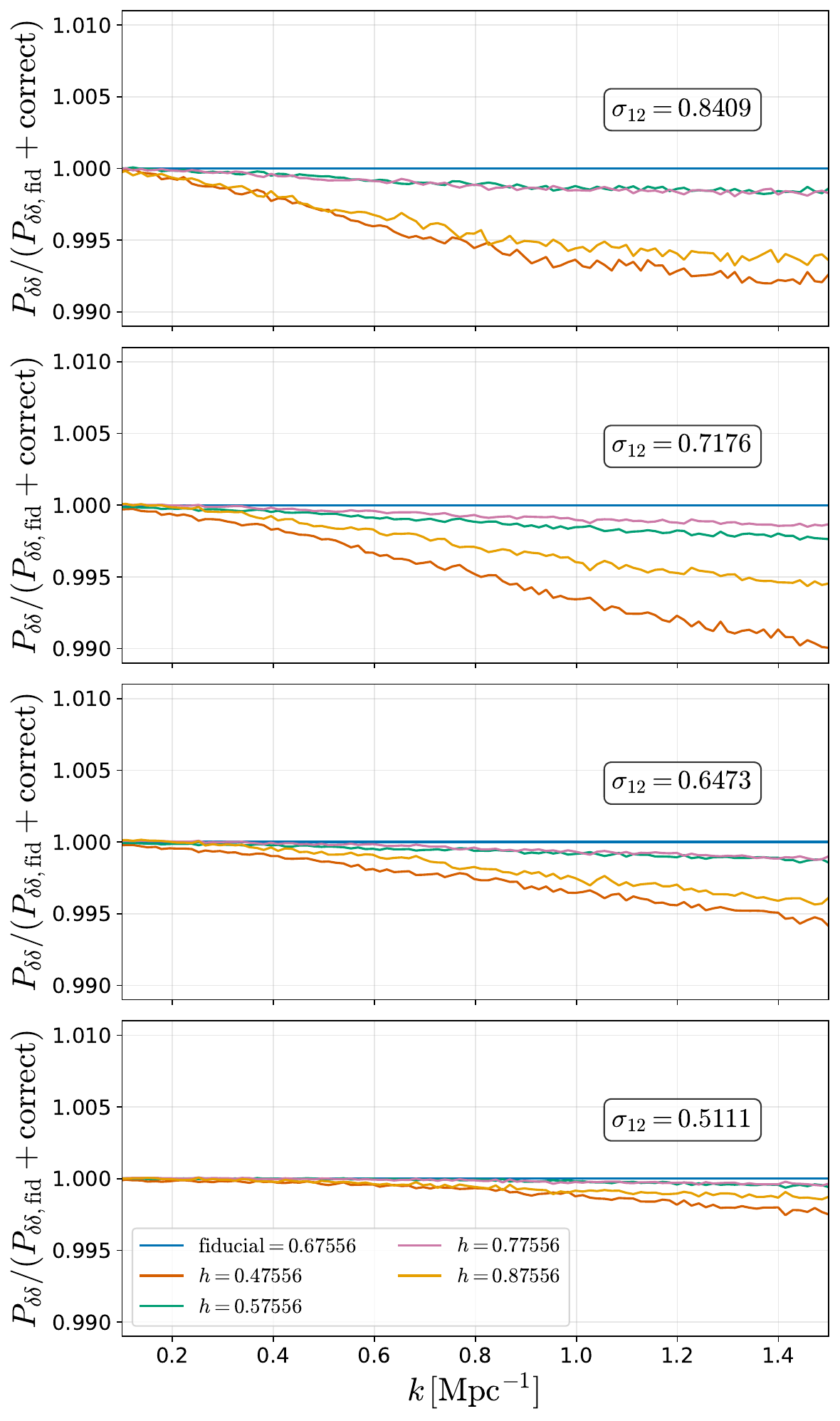}\\
    {\small (b) With nonlinear correction.}
\end{minipage}
    \caption{Validation of the evolution mapping approach for the density field, with (right panels) and without (left panels) the nonlinear correction of Eq.~\eqref{eq:h-nl-corr}. Note the different vertical scales used in the left- and right-hand panels.}   \label{fig:delta_panels} \vspace{12pt}
\end{figure}

In Fig.~\ref{fig:delta_panels} we first validate the evolution mapping approach for the density power spectrum. In the left panels we apply no nonlinear correction and just use Eq.~\eqref{eq:evol}, while in the right panels we apply the first-order correction given in Eq.~\eqref{eq:h-nl-corr}.
Note that, while for small values of $\si_{12}\sim 0.5$ the correction is not very significant, it improves the fit by more than a factor of 10 when $\si_{12}\sim 0.8$.  This result is in good agreement with the findings of~\cite{Esposito:2024qlo}.

\begin{figure}[htbp]
    \centering
\begin{minipage}{0.45\textwidth}
    \centering
    \includegraphics[width=\linewidth]{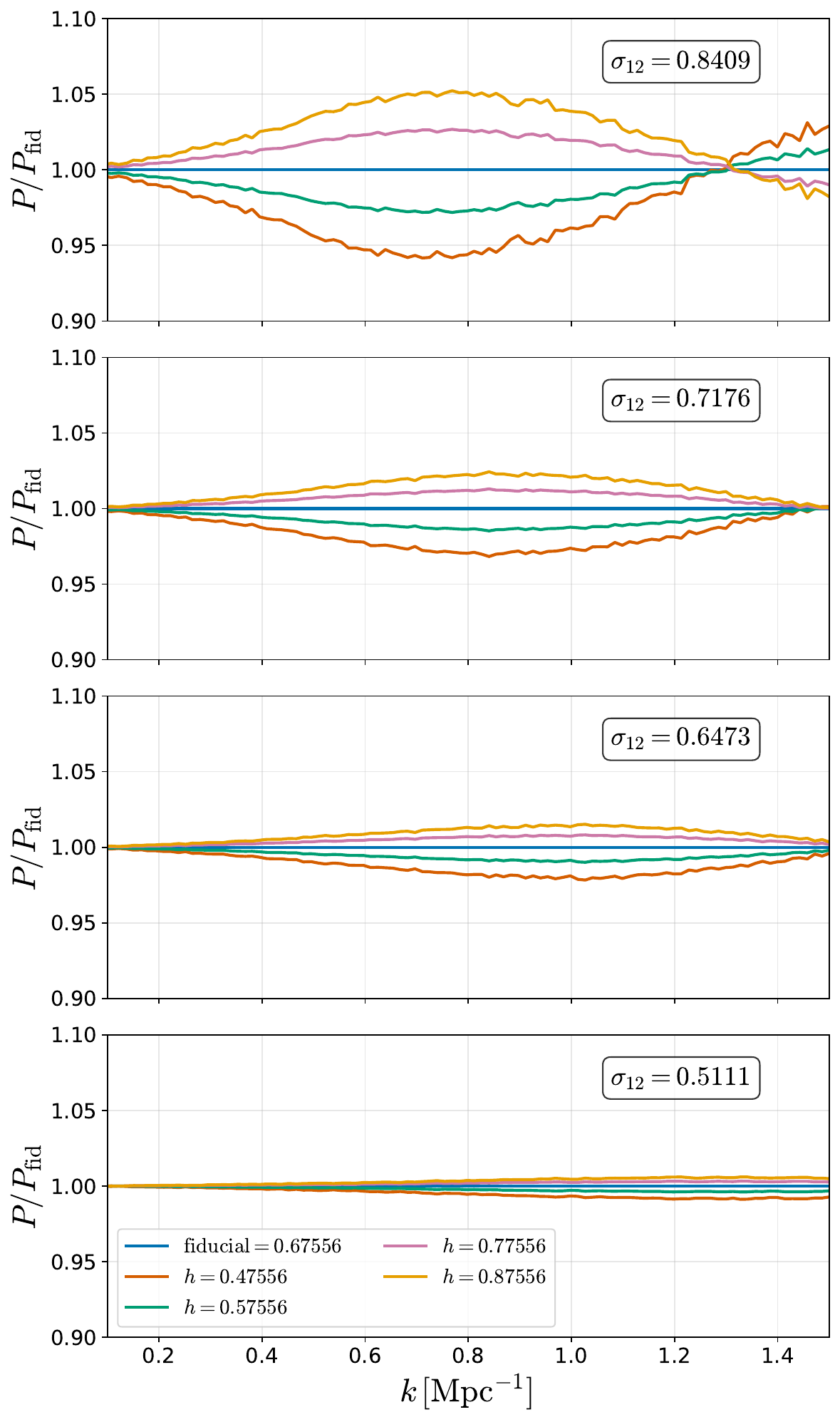}\\
    {\small (a) Without nonlinear correction.}
\end{minipage}
\hfill
\begin{minipage}{0.45\textwidth}
    \centering
    \includegraphics[width=\linewidth]{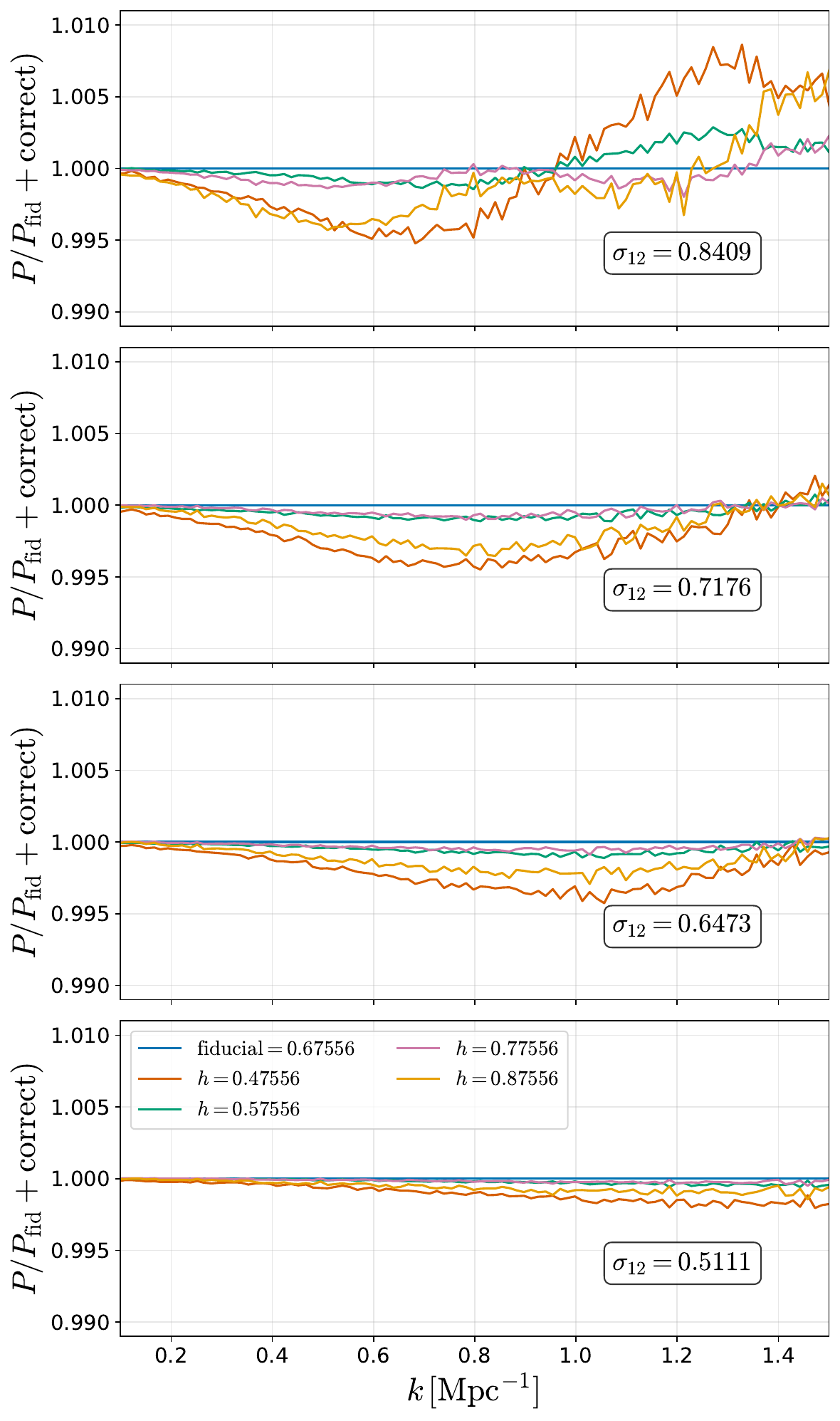}\\
    {\small (b) With nonlinear correction.}
\end{minipage}
    \caption{Validation of the evolution mapping approach for the velocity field, with (right panels) and without (left panels) the nonlinear correction of Eq.~\eqref{eq:h-nl-corr}. Note the different vertical scales used in the left- and right-hand panels.}\label{fig:theta_panels}
\end{figure}

In Fig.~\ref{fig:theta_panels}, we present analogous results for the velocity divergence power spectrum from our simulations. The estimation of the velocity power spectrum from simulations is not trivial as the velocity is best known where there are many particles and the density is high, while there is very little information in underdense regions. There are various ways to deal with this difficulty, e.g.,~via a Voronoi or a Delaunay tessellation~\citep{Pueblas:2008uv, Cautun:2011gf} or a phase-space interpolation technique~\citep{Hahn:2014lca}. Here we follow the prescription of~\cite{Jelic-Cizmek:2018gdp} where the velocity field in empty grid cells is approximated through linear growth from the time when there was still a particle in this cell (rescaled method). In~\cite{Jelic-Cizmek:2018gdp} it is shown that this relatively simple method is in good agreement with other, more sophisticated approaches, like the Delaunay tessellation.
We show the deviations from the linear evolution mapping, expressed by Eq.~\eqref{eq:evol}, in the nonlinear regime, for the velocity power spectra. Again the errors are significantly reduced from more than 5\% for the highest value of $\si_{12}$ to less than 0.8\% for all values of $h$ studied in our simulations, $0.47556\leq h\leq 0.87556$, exploring scales up to $k=1.5\,\mathrm{Mpc}^{-1}$.
Note that in each panel, we compare power spectra at fixed $\sigma_{12}$ and varying redshift, as reported in Table~\ref{table:h-sims}.

As $\sigma_{12}$ decreases, nonlinearities become typically less important and therefore the linear evolution mapping (i.e.\ without the nonlinear correction) becomes more accurate. We also note that while for the velocity spectra shown in Fig.\ \ref{fig:theta_panels} the linear evolution mapping is good to about 5\%,
for the density field the deviations exceed the $10\%$ level at high $\sigma_{12}$ but similarly improve at lower values.

\section{Beyond evolution mapping: the shape parameter}
\label{sec:shape-cdm}

Next, we study the dependence of the velocity power spectrum on a cosmological parameter that does not simply alter the amplitude of the power spectrum but also its shape, namely $\omega_{\rm cdm}$.
To this end, we consider the ratio of the nonlinear normalized velocity power spectra
\begin{equation}
   R_{\rm NL}(k, z,\om_{\rm cdm}) = \frac{P(k, z, h_{\rm fid},\om_{\rm cdm})}{P(k, z, h_{\rm fid},\om_{\rm cdm}^{\rm fid})} \, .
\end{equation} 
In this case, we compare power spectra at fixed redshift, rather than at fixed $\sigma_{12}$.
As a preliminary test, we perform a resolution test by running 
simulations for the fiducial cosmology and two test cosmologies at different resolutions. We fix the box size to 
$L_{\rm box} \simeq 190\,{\rm Mpc}$, as in the previous section, and vary the number of grid points per side: $N_{\rm grid} = 512$ for the low-resolution run and $N_{\rm grid} = 1024$ for the high-resolution run.

\begin{figure}[htbp]
    \centering
\includegraphics[width=0.7\linewidth]
{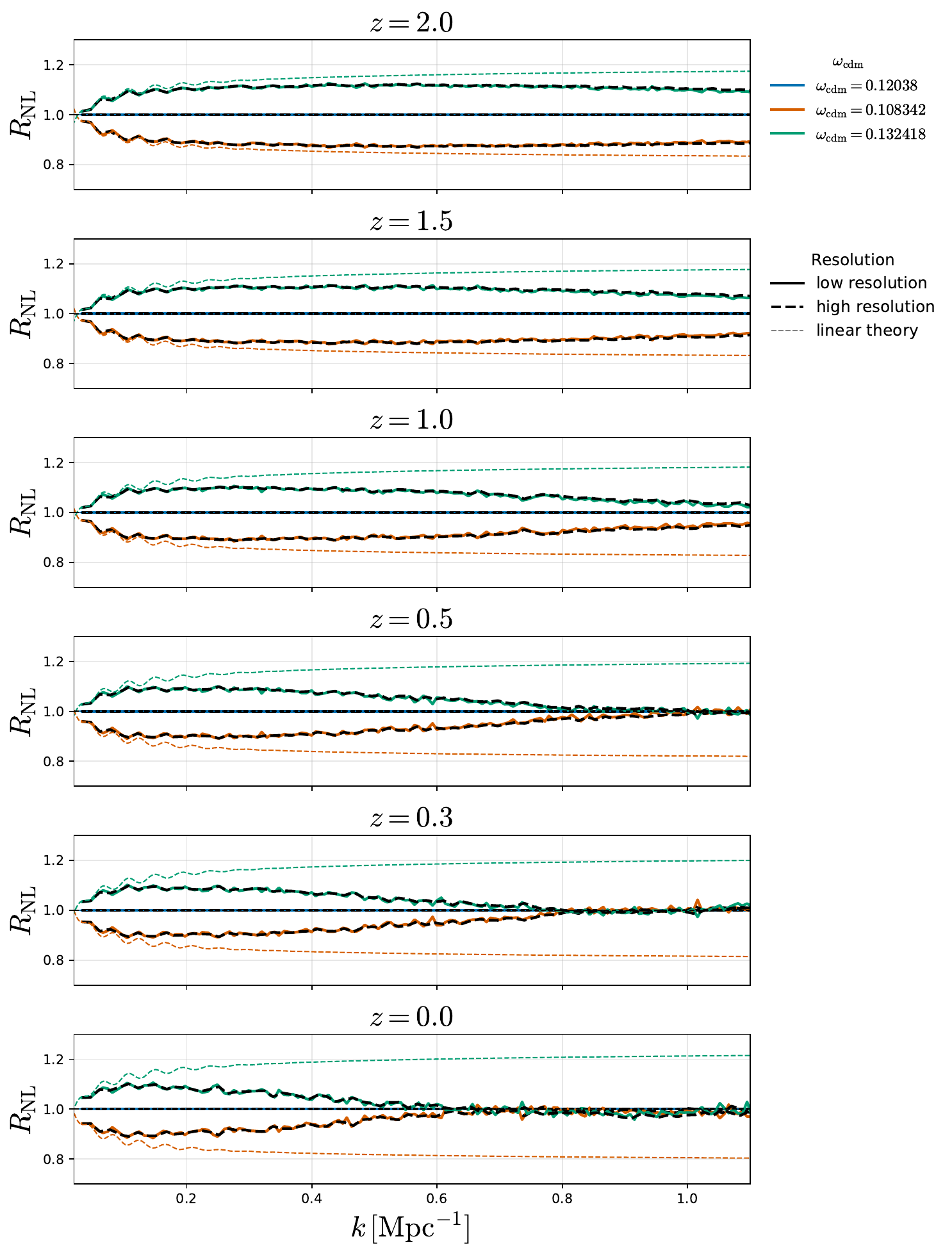}
    \caption{
    Ratio of the normalized velocity-divergence power spectrum,
$R_\mathrm{NL} = P/P_{\rm fid}$,
    for different values of $\omega_{\rm cdm}$ and redshifts.
    Colors denote the cosmological model, while line styles distinguish
    low-resolution simulations, high-resolution simulations, and linear theory. The nonlinear velocity power spectrum clearly deviates more from the linear spectrum for $k > 0.1 ~\mathrm{Mpc}^{-1}$ at lower redshifts. The numerical results for the two resolutions agree, indicating that the simulations are converged on the scales shown here.}
\label{fig:Pk_theta_ratio_omega_cdm}\vspace{12pt}
\end{figure}

Fig.~\ref{fig:Pk_theta_ratio_omega_cdm} shows the ratio $R_{\rm NL}$ for the result of this convergence test.  We find that our low-resolution runs are well converged for this quantity up to $k = 1\,{\rm Mpc}^{-1}$. 
Since the nonlinear ratios converge for our low-resolution runs, we run a more complete set of simulations sampling more values of $\omega_{\rm cdm}$ within this range.

On large scales, $k<0.1\,\mathrm{Mpc}^{-1}$ for $z=0$ and $k<0.2\,\mathrm{Mpc}^{-1}$ for $z=2$, the nonlinear ratios recover the linear-theory values, whereas scale-dependent deviations develop toward smaller scales as a consequence of nonlinear gravitational evolution.

Surprisingly, the nonlinear ratio of the power spectra is always closer to 1 than the linear one; and when the ratio predicted from linear theory becomes about 1.2, $R_{\rm NL}$ tends to unity. As is seen in Fig.~\ref{fig:Pk_theta_ratio_omega_cdm}, in our examples this happens roughly at $k=1/$Mpc for $z=0.5$, and $k=0.8/$Mpc for $z=0.3$ and at $k=0.6/$Mpc for  $z=0$.

In other words, in the `strongly nonlinear' regime, reached on small scales and low redshifts, the normalized velocity divergence power spectrum appears not to depend on $\om_{\rm cdm}$ any more. This means that when over-densities are high enough, the mean background density $\omega_\mathrm{cdm}$ no longer affects the peculiar velocity, which is then determined entirely by the density contrast as we shall see. This is interesting for simulations that want to resolve very small scales as they no longer have to care about the parameter $\om_{\rm cdm}$. Note, however that it is important to use  $\om_{\rm cdm}$ and not  $\Om_{\rm cdm}$ as the latter depends also on the evolution parameter $h$.

We now introduce a model for $R_{\rm NL}$ that captures the behavior described above. 
We seek a functional form that recovers the linear ratio at $k \ll k_{\rm NL}$ and approaches unity at $k \gg k_{\rm NL}$, where $k_{\rm NL}$ is a scale marking the onset of nonlinear effects, and where we assume that the ratio becomes insensitive to $\omega_{\rm cdm}$.
These requirements are satisfied by the functional form
\begin{equation}
R_{\rm NL} (k, z, \omega_{\rm cdm}) = 1 + W(k, k_{\rm NL}(z, \omega_\mathrm{cdm})) \left( R_{\rm lin} (k, z, \omega_{\rm cdm}) - 1 \right), 
\end{equation}
where $W(k, k_{\rm NL})$ is a filter function that approaches unity in the limit $k \ll k_{\rm NL}$ and vanishes in the limit $k \gg k_{\rm NL}$.
Here, $R_{\rm lin}(k, z, \omega_{\rm cdm})$ denotes the linear-theory counterpart of $R_{\rm NL}$, defined as
\begin{equation}
R_{\rm lin}(k, z, \omega_{\rm cdm}) = \frac{P_L(k, z, h_{\rm fid},\om_{\rm cdm})}{P_L(k, z, h_{\rm fid},\om_{\rm cdm}^{\rm fid})} \ ,
\end{equation}
indicated as dashed lines in Fig.~\ref{fig:Pk_theta_ratio_omega_cdm}.
We model the filter function as
\begin{equation}
W(k, k_{\rm NL}) = \frac{2}{1 + \exp\left(k / k_{\rm NL}\right)} \ ,
\end{equation}
and note that a simple exponential cutoff gives similar results.

Next, we need a model for the functional form of $ k_{\rm NL}$.
In a first step, we fit the ratio separately for different cosmologies and redshifts, in order to gain insight into how $k_{\rm NL}$ depends on $\om_{\rm cdm}$ and on redshift. Inspecting the best-fit values of $k_{\rm NL}$, we find that it can be modeled as a function of $\sigma_{12}$ alone, $k_{\rm NL} = k_{\rm NL}(\sigma_{12})$. This supports our hypothesis that once the density contrast is sufficiently high, it is all that matters for the velocity field, and the background matter density becomes unimportant.
We therefore perform a joint fit for all the data, assuming that $k_{\rm NL}$ follows a simple power law,
\begin{equation}
k_{\rm NL} =  k_0 \, \sigma_{12}^{-\alpha} \,.
\end{equation}
The resulting best-fit values are $k_0 = 0.139\,\mathrm{Mpc}^{-1}$ and $\alpha = 2.056$. At $z=0$, this nonlinearity scale becomes
$k_{\rm NL}(z=0) \simeq 0.2\,\mathrm{Mpc}^{-1}$. For $k=k_{\rm NL}$ our filter evaluates to $W(k_{\rm NL},k_{\rm NL}) \simeq 0.54$.
In the joint fit, we include scales up to $k = 1\,{\rm Mpc}^{-1}$, consistent with the convergence limit established by our resolution test.

\begin{figure}[htbp]
    \centering
\includegraphics[width=0.7\linewidth]
{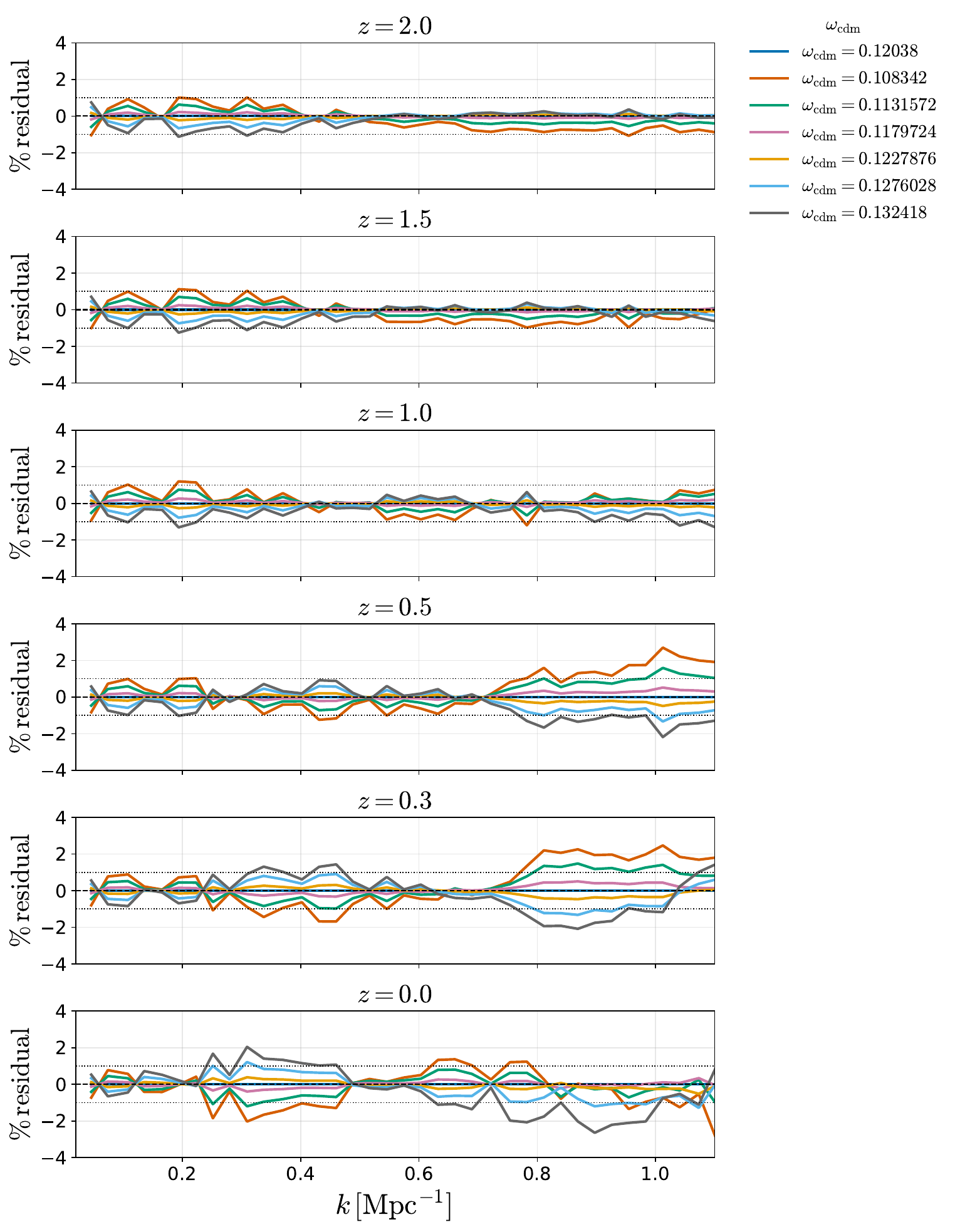}
\caption{Percentage residuals of the model for $R_{\rm NL}$, defined as $100 \times (R_{\rm NL}^{\rm sim} - R_{\rm NL}^{\rm model}) / R_{\rm NL}^{\rm sim}$, shown as a function of wavenumber $k$ at different redshifts. Different colors correspond to different values of $\omega_{\rm cdm}$, as indicated in the legend. The dashed black line marks a zero residual, and the dotted lines indicate the $\pm 1\%$ levels.}
\label{fig:residuals_omega_cdm}
\end{figure}

In Fig.~\ref{fig:residuals_omega_cdm} we show the percentage residuals of the model with respect to the simulations. The model achieves about $2\%$ accuracy across all redshifts and values of $\omega_{\rm cdm}$ explored in our simulation suite, up to $k = 1\,{\rm Mpc}^{-1}$.

\section{Full model for the velocity divergence power spectrum}
\label{sec:full-model}

In this section, we combine the two ingredients developed in Sections~\ref{sec:evol-h} and~\ref{sec:shape-cdm} into a complete model for the velocity power spectrum for arbitrary values of $\{\omega_{\rm cdm}, h\}$ in terms of the fiducial model. Using the evolution mapping approach, we write
\begin{align}
P_{\rm mod}(k, \sigma_{12}, \omega_{\rm cdm}, h) &= P(k, \sigma_{12}, \omega_{\rm cdm}, h_{\rm fid})
+ \frac{\partial P(k, \sigma_{12}, \omega_{\rm cdm}, h)}{\partial h}\big|_{h_{\rm fid}}
(h-h_{\rm fid}) \\
&= \frac{P(k, \sigma_{12}, \omega_{\rm cdm}, h_{\rm fid})}{ P(k, \sigma_{12}, \omega_{\rm cdm}^{\rm fid}, h_{\rm fid})}  P(k, \sigma_{12}, \omega_{\rm cdm}^{\rm fid}, h_{\rm fid}) + \frac{\partial P(k, \sigma_{12}, \omega_{\rm cdm}, h)}{\partial h}\big|_{h_{\rm fid}} 
(h-h_{\rm fid}) \\
&= R_{\rm NL}(k,z(\si_{12}),\omega_{\rm cdm})P(k, \sigma_{12}, \omega_{\rm cdm}^{\rm fid}, h_{\rm fid}) + \frac{\partial P(k, \sigma_{12}, \omega_{\rm cdm}, h)}{\partial h}\big|_{h_{\rm fid}} 
(h-h_{\rm fid})\,, \\
&\approx R_{\rm NL}(k,z(\si_{12}),\omega_{\rm cdm})
\left[P(k, \sigma_{12}, \omega_{\rm cdm}^{\rm fid}, h_{\rm fid}) + \frac{\partial P(k, \sigma_{12}, \omega_{\rm cdm}^{\rm fid}, h)}{\partial h}\big|_{h_{\rm fid}} 
(h-h_{\rm fid})\right]\,, \label{e:full-model}
\end{align}
where $R_{\rm NL}$ is the nonlinear ratio modeled in Section~\ref{sec:shape-cdm} that is now evaluated at the redshift corresponding to a fixed $\si_{12}$, and the $h$-derivative term is the correction developed in Section~\ref{sec:evol-h}.
In the last line of Eq.~\eqref{e:full-model}, we assume that the dependence on $\omega_{\rm cdm}$ factorizes from the first-order correction in $h$, such that the derivative at a generic value of $\omega_{\rm cdm}$ can be
approximated by rescaling the derivative measured in the fiducial cosmology by $R_{\rm NL}$.
To employ this formula in practice, we need to run about five $N$-body simulations to evaluate $ P(k, \sigma_{12}, \omega_{\rm cdm}^{\rm fid}, h_{\rm fid})$ and its $h$-derivative term. 

\begin{figure}[htbp]
    \centering
\includegraphics[width=0.8\linewidth]
{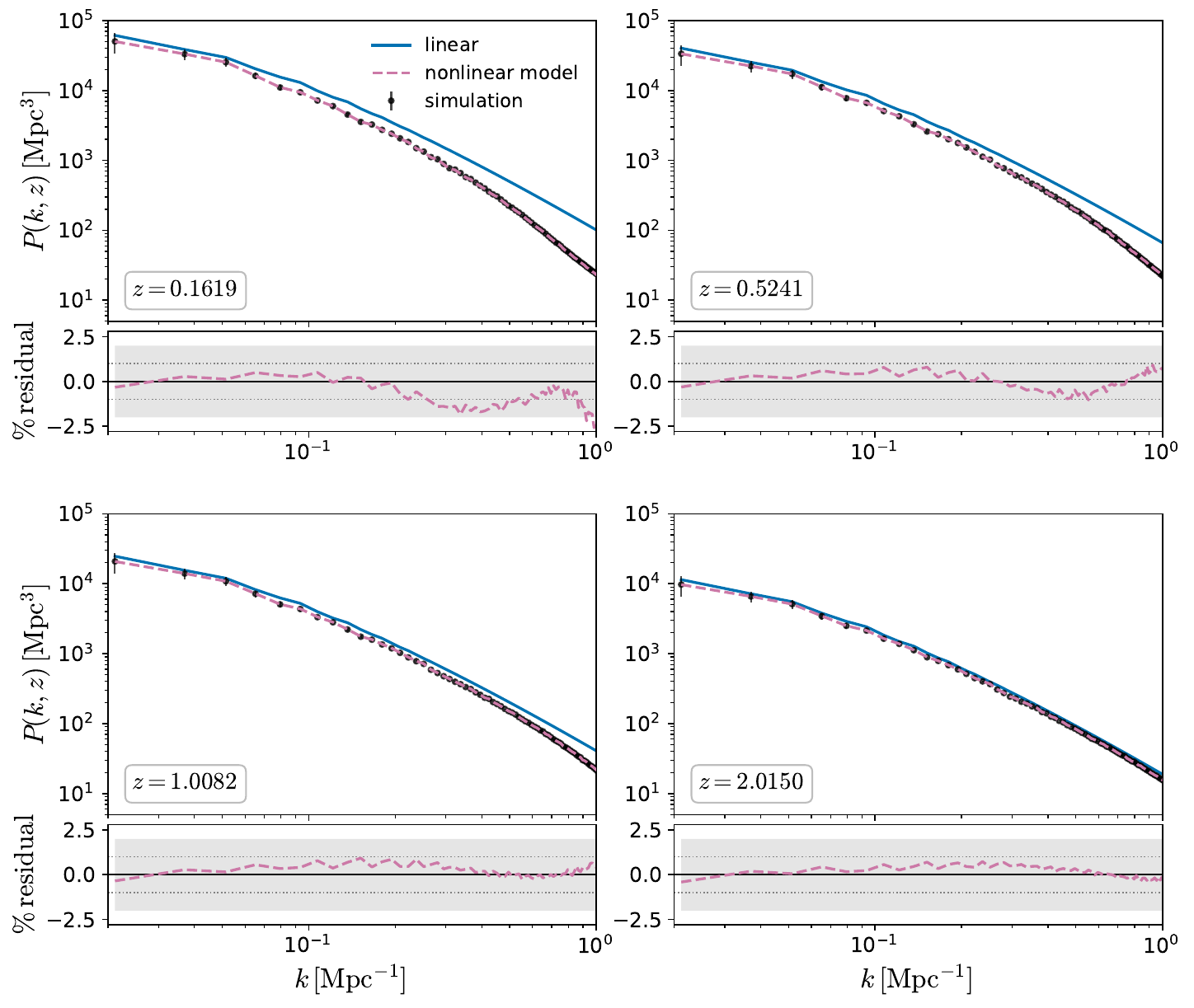}
    \caption{
Validation of the full nonlinear velocity-divergence power-spectrum model of Eq.~\eqref{e:full-model} against an independent simulation with $h=0.6$ and $\omega_{\rm cdm}=0.13$. The four panels show
    the rescaled velocity power spectrum $P(k,z)$ at
    $z=0.1619$, $0.5241$, $1.0082$, and $2.0150$. The solid blue curves show
    the linear prediction, the dash-dotted pink curves show our nonlinear
    model, and the black points with error bars show the simulation
    measurements. The lower sub-panels display the fractional residual. The gray bands indicate
    deviations within $\pm2\%$, while the dotted horizontal lines mark
    $\pm1\%$.
   }
\label{fig:Pk_theta_validation}\vspace{12pt}
\end{figure}

In Fig.~\ref{fig:Pk_theta_validation}, we validate the joint model against an independent test simulation with $h=0.6$ and $\omega_{\rm cdm}=0.13$, which was not used in its calibration. 
We perform the comparison at $z = 0.1619, 0.5241, 1.0082,$ and $2.0150$, chosen such that the test cosmology has the same $\sigma_{12}$ as the fiducial cosmology at $z = 0, 0.5, 1,$ and $2$, respectively. This choice allows us to compare the test simulation directly with the model, which is constructed using the fiducial simulation, at matched values of $\sigma_{12}$ without requiring any interpolation in redshift, thereby making the validation more accurate.
The linear prediction increasingly overestimates the velocity-divergence power on
nonlinear scales, whereas the full model closely follows the simulation measurements over the entire range $k\leq1\,\mathrm{Mpc}^{-1}$. The maximum absolute residual is approximately $2.5\%$ at the lowest redshift shown, $z=0.1619$, and decreases to approximately $1.0\%$, $0.9\%$, and $0.7\%$ at $z=0.5241$, $1.0082$, and $2.0150$, respectively. Thus, the model
reproduces the full $N$-body result at the percent level for
$z\gtrsim0.5$ and remains accurate to within approximately $2.5\%$ at the lowest redshift considered.

As already mentioned, we do not attempt to also model $\omega_{\rm b}$ as this quantity is very precisely measured by the CMB.

A Python implementation of the velocity-divergence power-spectrum model,
together with an example notebook and the data required to reproduce the validation presented in this work, is publicly available at
\url{https://github.com/leporif/velocity_power_spectrum}.
The code relies on the code {\sc class}~\citep{Blas:2011rf, Lesgourgues:2011re} for most of the cosmological computations.

\section{Conclusions}
\label{sec:concl}
In this paper we use the evolution mapping formalism to derive an approximation for the velocity divergence power spectrum, $P_{\th\th}$, in a $\La$CDM cosmology at fixed $\om_{\rm b}$ and $n_{\rm s}$ for  physically sensible but wide ranges of values for $h$ and $\om_{\rm cdm}$. 
Our model reproduces the simulation results to within $2.5\%$ at $z\simeq 0$ and to sub-percent accuracy for $z \geq 0.5$, over the full range $k < 1\,{\rm Mpc}^{-1}$ and for all parameter values considered in this work.

To set up the model, five simulations at the fiducial value of $\om_{\rm cdm}$ are required to determine the fiducial spectrum and estimate its derivative with respect to $h$. 
The fact that five simulations are still needed here is more of an aesthetic limitation than a practical drawback.
A possible replacement of $P(k,\si_{12}, \om^{\rm fid}_{\rm cdm}, h_{\rm fid})$ and its derivative with respect to $h$ with an analytical fitting formula is left for future work.

In Appendix~\ref{ap:vort}, we also present a simple model for the vorticity power spectrum which has been shown to be subdominant on the scales $k<1\,\mathrm{Mpc}^{-1}$ that we have investigated here.
Furthermore, in Appendix~\ref{app} we show that for a statistically homogeneous and isotropic velocity field, measuring its radial component is actually sufficient to estimate both the divergence and the vorticity power spectrum or correlation function.

These results can be used to model redshift-space distortions, redshift perturbations from supernova measurements or directly the velocity power spectrum and correlation function for large peculiar velocity surveys like Cosmicflows-4 \citep{Tully:2022rbj} or similar.

On the repository \url{https://github.com/leporif/velocity_power_spectrum} one can find the velocity power spectrum for our fiducial simulation and its $h$-derivative ready to use.
More generally, the code can be
adapted to other simulation sets by supplying the power spectrum of an
arbitrary fiducial simulation and four additional simulations at nearby
values of $h$, from which the $h$-derivative is estimated. It can then
generate a surrogate prediction for the nonlinear velocity-divergence power
spectrum of any cosmology within the calibrated parameter range, with an
accuracy of approximately $2.5\%$ at $z\simeq0$ and sub-percent accuracy for
$z\geq0.5$ over the range $k<1\,{\rm Mpc}^{-1}$.

\section*{Acknowledgments}
The authors acknowledge financial support from the Swiss National Science Foundation (SNSF). FS acknowledges support from SNSF through the Postdoc.Mobility fellowship (grant no. P500-2\_235508).
Some of the computations underlying this work were performed on the Baobab cluster at the University of Geneva. This work uses data originally generated by a grant from the Swiss National Supercomputing Centre (CSCS) under project ID s710.

\appendix

\section{Vorticity}
\label{ap:vort}

The vorticity contribution is modeled separately from the velocity divergence. Since the simulations used to calibrate the divergence model do not have sufficient spatial resolution to obtain a converged vorticity power spectrum, we adopt instead a phenomenological broken power-law model with a plateau, calibrated to reproduce the measurements of \cite{Jelic-Cizmek:2018gdp}:

\begin{equation}
P_{\om\omega}(k,z)=A(z)
\frac{2\,\left(k/k_{\rm p}\right)^a}
{\left[1+\left(k/k_{\rm p}\right)^b\right]
\left[1+\left(k/k_{\rm s}\right)^b\right]^{d/b}}.
\end{equation}

\begin{equation}
A(z)=P_0 D_1^7(z),
\qquad
k_{\rm p} =h_{\rm fid}(1+z)\,[{\rm Mpc}^{-1}],
\qquad
k_{\rm s}=\alpha k_{\rm p}.
\end{equation}
We adopt the parameter values $a=2.5$, $b=2.5$,
$d=1.5$, $\alpha=1.5$, $P_0=5/h^3_{\rm fid}\,[{\rm Mpc}^{-3}]$.

The dependence of this prescription on the cosmological parameters, in particular $h$ and on $\omega_{\rm cdm}$, has not been calibrated against dedicated simulations and should therefore be regarded as approximate.

\begin{figure}[htbp]
    \centering
\includegraphics[width=0.8\linewidth]
{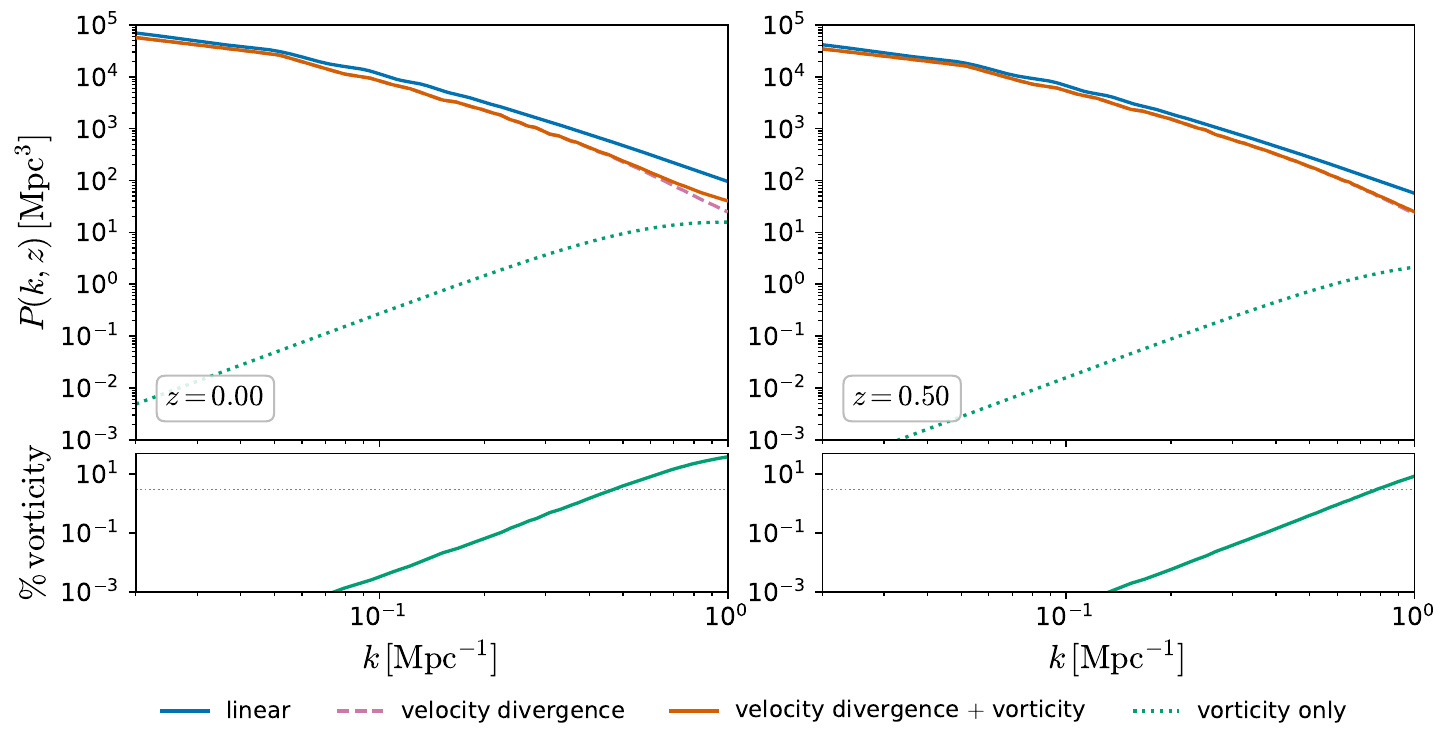}
    \caption{
Contribution of vorticity to the velocity power spectrum at
$z=0$ and $z=0.5$. The upper panels compare the linear velocity-divergence
power spectrum with the nonlinear model evaluated with and without the vorticity contribution; the vorticity power spectrum is also shown separately. The lower panels display the fractional vorticity contribution in percentage. The horizontal dotted line
indicates the $3\%$ level.
   }
\label{fig:Pk_vorticity}\vspace{12pt}
\end{figure}

Fig.~\ref{fig:Pk_vorticity} illustrates the size of this term in the fiducial cosmology at
$z = 0$ and $z = 0.5$.
The vorticity contribution is entirely negligible on large scales and grows steeply toward small
scales, where the plateau imposed by $k_{\rm s}$ limits its rise. At $z = 0$ it reaches
$\sim 40\%$ of the velocity-divergence spectrum at $k = 1\,{\rm Mpc}^{-1}$, the edge of the range over which our model is calibrated, and it stays below $\sim 3\%$ for $k \lesssim 0.5\,{\rm Mpc}^{-1}$. The steep growth-factor scaling suppresses the term rapidly towards higher redshift. Already at $z = 0.5$ it drops to $\sim 10\%$ at
$k = 1\,{\rm Mpc}^{-1}$, and at $z \gtrsim 1$ it is negligible over the whole range we model. Vorticity is therefore relevant only at low redshift and on the smallest scales we model, which justifies treating it as an additive correction  to the divergence power spectrum evaluated at the fiducial parameter values rather than including it in the evolution-mapping calibration.

\section{Measuring the gradient and curl components of the velocity power spectrum}\label{app}
In this Appendix we show that, when assuming statistical homogeneity and isotropy, measuring the radial velocity field is in principle sufficient to determine both, the vorticity and the gradient power spectrum or correlation function.

We assume we have measured the radial velocity field via some redshift and distance measurements,
\be
\frac{v_r}{c} =\frac{\de z}{1+z} \sim H(z)(d_{\rm measured}-d_{\rm model}(z)) \,.
\ee

In general, $v_r$ does certainly not determine the full velocity field $\bv$. However, if we know that the velocity field is a gradient,
\be
\bv(\bx) = \bnabla v(\bx)\,,
\ee
then from the radial velocity, $v_r=\dd_rv$ we can in principle determine 
$$
v(\bx) = \int_0^d v_r(r\hat\bx)dr
$$
where $\hat\bx=\bx/d =\bn$ is the unit vector in direction $\bx$ and $d=|\bx|$. From the velocity potential $v$ we then obtain the velocity simply by  taking the gradient.

\subsection{Splitting a vector field correlator into gradient and curl}
However, if $\bv$ also contains a curl component, i.e., a vorticity
\be
\bv(\bx) = \bnabla v(\bx) +\bnabla\times\ba(\bx)\,,
\ee
this no longer works. From $v_r$ we cannot infer $v$ and $\ba$ without additional information.

We now show that assuming a stochastic velocity field that is statistically homogeneous and isotropic, we can actually estimate the spectra of both $v$ and $\ba$ from the measurement of the correlation function of $v_r$ alone.

To see this we introduce the velocity power spectrum which contains a gradient and a curl part,
\be\label{e:vspec}
\langle v_i(\bk)v_j^*(\bk')\rangle = (2\pi)^3\de(\bk-\bk')\left[\hat k_i\hat k_j P_G(k) +(\de_{ij}-\hat k_i\hat k_j) P_C(k)  \right] \,.
\ee
Here $k=|\bk|$ and $\hat\bk=\bk/k$. The tensorial structure of the power spectrum is determined by statistical homogeneity and isotropy (and parity invariance, otherwise a term $i\epsilon_{ijm}\hat k_mP_H(k)$ with negative parity would also be possible).

The first term in \eqref{e:vspec} is pure gradient while the second one is pure curl. Therefore, we can also introduce the divergence and the vorticity of the velocity field,
\begin{align}
\theta &= \bnabla\cd\bv\,, \qquad  &\langle \theta(\bk)\theta^*(\bk')\rangle &=(2\pi)^3\de(\bk-\bk')P_{\th\theta}(k)\,, \quad  &P_{\th\theta}(k) &= k^2P_G(k) \\
\om_i &= (\bnabla\times\bv)_i \,,  \quad &\langle \om_i(\bk)\om_j^*(\bk')\rangle &= (2\pi)^3\de(\bk-\bk')(\de_{ij}-\hat k_i\hat k_j)P_{\om\om}(k)\,, \quad  &P_{\om\om}(k) &= k^2P_C(k)   \,.
\end{align}

Note that $P_{\th\om_j}=0$, as it would define a preferred direction. But also $P_{v_i\om_j}=0$ if parity invariance is respected as we assume here, $P_H\equiv 0$. 

Fourier transforming \eqref{e:vspec}, with $\br=\bx-\by$ and $r=|\br|$, we obtain 

\be \label{e:2pfvspec}
\xi_{ij}(\br) = \langle v_i(\bx)v_j(\by)\rangle = \dd_i\dd_jf(r) +(\de_{ij}\De-\dd_i\dd_j)g(r) \,,
\ee

where $ \De = \de^{ij}\dd_i\dd_j$,

\bea
f(r) &=& \frac{-1}{2\pi^2}\int_0^\infty dk P_G(k)j_0(kr) \qquad   \mbox {and}  \label{e:f}\\
g(r) &=& \frac{-1}{2\pi^2}\int_0^\infty dk P_C(k)j_0(kr) \, . \label{e:g}
\eea

Using

\bea
 \dd_i\dd_jf(r)  &=& \hat r_i\hat r_jf'' -\frac 1r (\hat r_i\hat r_j-\de_{ij})f'  \qquad   \mbox {and}\\
 (\de_{ij}\De-\dd_i\dd_j)g(r)   &=& ( \de_{ij}-\hat r_i\hat r_j)g''  +\frac 1r (\hat r_i\hat r_j+ \de_{ij})g' \,,
\eea

we can write the correlation function in the form

\be\label{e:xiij}
\xi_{ij}(\br) =  \hat r_i\hat r_j(f''-g'') + \de_{ij}g''  -\frac 1r (\hat r_i\hat r_j-\de_{ij})(f'-g') + \frac 2r  \de_{ij}g' \,.
\ee

We assume that we can measure $v_r(\bx) = \bv(\bx)\hat\bx \equiv  \bv(d\bn)\bn$ in arbitrary directions $\bn$ and 
for arbitrary distances $d=|\bx|$, so that we can estimate its correlation function, 
\be
\label{e:2pfvr}
\langle v_r(\bx) v_r(\bx')\rangle =  \xi_{ij}(\br)n^in^{'j}
\ee
where $\br = \bx-\bx'$, $\bn'=\bx'/d'$ and $d'=|\bx'|$. Defining
\be
r= |\br|=\sqrt{d^2+d'^2-2dd'\mu} \,, \qquad \mu=\bn\cd\bn' \,, 
\ee
and combining Eq.~\eqref{e:xiij} and Eq.~\eqref{e:2pfvr}, we obtain
\bea
\langle v_r(\bx) v_r(\bx')\rangle &=& \frac 1{ r^2}\left[(d^2+{d'}^2)\mu-dd'(1+\mu^2) \right](f''-g'') +\mu g''+\frac 2 r \mu g'  \nonumber \\
    &&  -\frac 1{ r^3}\left[(d^2+{d'}^2)\mu -dd'(1+\mu^2) -r^2\mu\right](f'-g') \,,   \label{e:corv}
\eea
This is a function of $d,~d'$ and $-1\leq\mu\leq 1$. Let us first evaluate it for $\mu=0$, i.e., $\bx$ and $\bx'$ orthogonal. For this case, the terms involving $g$ alone drop and we obtain an equation for $f-g$.
Denoting $\langle v_r(\bx) v_r(\bx')\rangle$ for orthogonal  $\bx$ and $\bx'$ as  $\xi_\perp(d,d')$ we have
\be
\boxed{\frac{d^2+ {d'}^2}{dd'}\xi_\perp(d,d') = \frac{1}{\sqrt{d^2+ {d'}^2}}(f'-g') -(f''-g'') \,.}
\ee
This differential equation can be solved for $f-g$ which is a function of $r=\sqrt{d^2+{d'}^2}$ only. Hence also  $\xi_\perp(d,d')/(dd')$ is a function of $r$.  This is true for the following reason: if $\bn\perp \bn'$, $\langle v_r(\bx) v_r(\bx')\rangle$ given in \eqref{e:corv} is a function of $r$, times $(\hat\br\cd\bn)(\hat\br\cd\bn') = (d- \mu d')(d\mu-d')/r^2=-dd'/r^2$   where the last equal sign holds since $\mu=0$.  Therefore   $\xi_\perp(d,d')/(dd')$ is a function of $r$ alone. The boundary condition for the solution $f-g$ should be such that $(f-g)(r)\ra 0$ for $r\ra\infty$. Note that, since $\dd_dr = d/r$ and $\dd_{d'}r = d'/r$  when $\mu=0$, we can also write
\be
\boxed{\xi_\perp(d,d') = -\dd_d\dd_{d'}(f-g)(r) \,, }
\ee
Also from this expression we can see that $\xi_\perp(d,d') /(dd')$ is a function of $r$ alone.

Once we have the solution for $f-g$, we can measure the correlation function \eqref{e:corv} for arbitrary values of $\mu$ to obtain a differential equation for $g$,
\bea
\langle v_r(\bx) v_r(\bx')\rangle - \frac 1{ r^2}\left[(d^2+{d'}^2)\mu - dd'(1+\mu^2)\right](f''-g'')  \hspace*{0.5cm}   \nonumber\\  
   +\frac 1{ r^3}\left[(d^2+{d'}^2)\mu -dd'(1+\mu^2) -r^2\mu\right](f'-g')  \nonumber \\
=\mu g''+\frac 2 r \mu g' 
 \,.   \label{e:corv2}
\eea
Equivalently
\be  \label{e:corv3}
\boxed{\langle v_r(\bx) v_r(\bx')\rangle +\dd_d\dd_{d'}(f-g) = \mu( g''+\frac 2 r g') \,. }
\ee 
Solving this equation for arbitrary non-zero $\mu$, e.g.,~for $\mu=-1$, i.e., $r=d+d'$, yields $g$ and hence $f=(f-g)+g$.

The simplifying assumption made here was that we know $\langle v_r(\bx) v_r(\bx')\rangle$ in all of space. We have neglected that observations can only be made on our past light cone, i.e.,\ for large $\bx$ we cannot neglect time dependence and retardation effects. But for sufficiently small scales, $k>1$ $h\,\mathrm{Mpc}^{-1}$ and low redshifts, where vorticity really matters, see~\cite{Jelic-Cizmek:2018gdp}, this should not be a very serious obstacle.

\subsection{Relating \texorpdfstring{$P_G$}{PG} and \texorpdfstring{$P_C$}{PC} to observables}
We have seen that  the observable perpendicular correlation function, $\xi_\perp$, relates to $f-g$ via
$$
\frac{r}{dd'}\xi_\perp(d,d') = \frac{f'-g'}{r^2}-  \frac{f''-g''}{r} = -\left[\frac{f'-g'}{r}\right]' \,. $$
The left-hand side (lhs) is an observable depending only on $r=\sqrt{d^2+{d'}^2}$.
To relate the right-hand side (rhs) to the power spectra $P_G$ and $P_C$ we use that, since $j_0' =-j_1$, from Eqs.~\eqref{e:f} and~\eqref{e:g} we have that
\be
\frac{f'-g'}{r} = \frac{1}{2\pi^2r}\int dk k \left(P_G(k)-P_C(k)\right)j_1(kr) \,.
\ee

Taking the derivative wrt $r$ on both sides and using some spherical Bessel function identities we obtain
\be
-\left[\frac{f'-g'}{r}\right]' = \frac{1}{2\pi^2r}\int dk k^2 \left(P_G(k)-P_C(k)\right)j_2(kr) \,.
\ee
Using the orthogonality relation of spherical Bessel functions,
\be
\int dy y^2j_n(ay)j_n(by) =\frac{\pi}{2 a^2}\de(a-b)
\ee
and multiplying above with $r^2j_2(k'r)$, we can isolate $P_G-P_C$ with the final result
\be
\boxed{P_G(k)-P_C(k)=4\pi\int_0^\infty \frac{dr r^4}{dd'}\xi_\perp(d,d') j_2(kr)}\,.
\ee

To find $P_C$ we now assume that we have measured the full radial correlation function~\eqref{e:2pfvspec} as a function of $d, d', \mu$ and $\br=\bx-\bx' =d\bn-d'\bn'$:
\bea
\langle v_r(\bx)v_r(\bx')\rangle &=& \xi_{ij}(\br)n^i{n'}^j =   \nonumber\\
&&  \hspace{-3cm} \frac{1}{(2\pi)^3}\int d^3k\left[(\hat\bk\cd\bn)(\hat\bk\cd\bn')\left(P_G(k)-P_C(k)\right)e^{i\bk\cd(d\bn-d'\bn')} +(\bn\cd\bn')P_C(k)e^{i\bk\cd(d\bn-d'\bn')}\right]  ~=  \nonumber\\
&& \hspace{-3cm}  \frac{1}{(2\pi)^3}\left[\int \frac{d^3k}{k^2}\left(P_G(k)-P_C(k)\right)\dd_d\dd_{d'}e^{i\bk\cd(d\bn-d'\bn')} +(\bn\cd\bn')\int d^3kP_C(k)e^{i\bk\cd(d\bn-d'\bn')}\right] \,.    \label{e:PCobs}
\eea
Using that the first integral just reproduces $f-g$ (see Eqs. (\ref{e:f},\ref{e:g})), which has already been related to $\xi_\perp$, we find for $(\bn\cd\bn')\neq 0$
\be
\frac{\langle v_r(\bx)v_r(\bx')\rangle+\dd_d\dd_{d'}(f-g)(r)}{(\bn\cd\bn')} =  \frac{1}{(2\pi)^3}\int d^3kP_C(k)e^{i\bk\cd(d\bn-d'\bn')} = \frac{1}{2\pi^2}\int_0^\infty dkk^2P_C(k)j_0(kr) \,.    \label{e:PCobs2}
\ee
Also here one can show directly that like the rhs, the lhs is a function of $r$ alone, e.g.,~by considering \eqref{e:corv3}. This can be used as a consistency check in the observations.
Comparison of \eqref{e:PCobs2} and  \eqref{e:corv3} also yields
\be
\frac{1}{r^2}\left(r^2g'\right)' = \frac{1}{2\pi^2}\int_0^\infty dkk^2P_C(k) j_0(kr) \,.
\ee
This can also be obtained directly from \eqref{e:g} using some spherical Bessel function identities.
Using again the orthogonality relation of spherical Bessel functions, we obtain
\begin{align}
P_C(k) &= 4\pi\int dr\left(r^2g'\right)'j_0(kr)  \\
 P_C(k) &=   4\pi\int dr r^2\frac{\langle v_r(\bx)v_r(\bx')\rangle+\dd_d\dd_{d'}(f-g)(r)}{(\bn\cd\bn')}j_0(kr)  \,.     \label{e:Pc}
\end{align}

We have seen that considering first the correlation function in directions that are $90^\circ$ apart allows us to determine the combination $f-g$ or $P_G-P_C$.
Using this function then for arbitrary directions with $\bn\cd\bn'\neq 0$ further allows us to isolate $g$ or $P_C$ and so to fully determine the 2-point correlation function and the power spectrum of the velocity field.\\
Note that $\dd_d\dd_{d'}(f-g)(r)$ in \eqref{e:Pc} cannot be replaced by $\xi_\perp$ since it equals $\xi_\perp/(dd')$ only if $\bx \perp \bx'$.

\bibliographystyle{apsrev4-1}

\bibliography{mybib}

\end{document}